\documentclass[aps,prl,reprint,superscriptaddress, showpacs]{revtex4-2}

\usepackage{graphicx} 
\usepackage{color} 
\usepackage{amsmath}

\begin{document}


\title{Improved limits on the spin- and velocity-dependent exotic interaction in the micrometer range}

\author{Sumin Li}
\thanks{These authors have contributed equally to this work.}
\affiliation{National Gravitation Laboratory, MOE Key Laboratory of Fundamental Physical Quantities Measurement, and School of Physics, Huazhong University of Science and Technology, Wuhan 430074, People's Republic of China}

\author{Wenbo Zhang}
\thanks{These authors have contributed equally to this work.}
\affiliation{National Gravitation Laboratory, MOE Key Laboratory of Fundamental Physical Quantities Measurement, and School of Physics, Huazhong University of Science and Technology, Wuhan 430074, People's Republic of China}

\author{Rui Luo}
\affiliation{National Gravitation Laboratory, MOE Key Laboratory of Fundamental Physical Quantities Measurement, and School of Physics, Huazhong University of Science and Technology, Wuhan 430074, People's Republic of China}

\author{Jinquan Liu}
\affiliation{National Gravitation Laboratory, MOE Key Laboratory of Fundamental Physical Quantities Measurement, and School of Physics, Huazhong University of Science and Technology, Wuhan 430074, People's Republic of China}

\author{Pengshun Luo}
\email[Corresponding author: ]{pluo2009@hust.edu.cn}
\affiliation{National Gravitation Laboratory, MOE Key Laboratory of Fundamental Physical Quantities Measurement, and School of Physics, Huazhong University of Science and Technology, Wuhan 430074, People's Republic of China}


\date{\today}

\begin{abstract}
Searching for the exotic interactions beyond the Standard Model of particle physics may solve some of the current puzzles in physics. Here the authors experimentally explore a spin- and velocity-dependent exotic interaction between the nucleons in a gold sphere and the electrons in a spin source in the micrometer range. The microfabricated spin source provides periodically varying spin density of electrons, resulting in a periodic exotic field. A cantilever glued with a gold sphere is used to measured the force acting on the gold sphere by the spin source. The spin source is driven to oscillate, and then the imaginary part of the signal is extracted at the 10th harmonic of the oscillation frequency, which effectively separates the exotic interaction from the spurious forces commonly present in such short-range measurements.  No signal of the exotic interaction is observed, then new limits on the coupling constant are set in an interaction range below 10 $\mu$m, with $f_{4+5} \le 2.2\times 10^{-9}$ at 2.1 $\mu$m.
\end{abstract} 

\maketitle
The Standard Model of particle physics (SM) stands as one of the most successful theories, describing the known fundamental particles and interactions other than gravity. Despite its triumphs, there remain several phenomena that lie beyond the scope of the SM, hinting at the existence of additional forces or particles that may weakly interact with the known constituents of matter. Some bosons, including axions\cite{Peccei1977, Weinberg1978, Wilczek1978}, majorons\cite{Chang1985}, familons\cite{Gelmini1983, Wilczek1982}, paraphotons\cite{Holdom1986, Appelquist2003}, and $Z'$ bosons\cite{Fayet1986, Langacker2009}, have been proposed in various extensions of SM. Some of them may be candidates for dark matter\cite{Rosenberg2000, Kim2010, Marsh2016, Chiles2022, An2023, Cline2024}, so it is particularly interesting and important to search for them. These hypothetical bosons may mediate exotic interactions\cite{Moody1984, Bogdan2006, Fadeev2019}, making it possible to search for them through the so-called fifth-force experiments\cite{cong2024}. In the general framework of quantum field theory, various interaction potentials are derived and classified according to either the spin-momentum forms or the types of physical couplings\cite{Bogdan2006, Fadeev2019, cong2024}. 

\begin{figure*}
	\centering
	\includegraphics {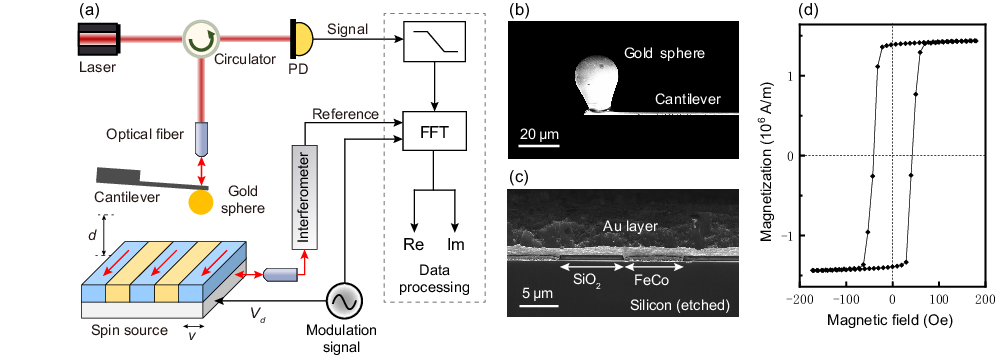}
	\caption{(a) Schematic drawing of the experiment. (b) Scanning electron microscopy (SEM) image of the cantilever and gold sphere. (c) SEM image of a spin source cross-section before glued to a glass support. (d) The magnetic hysteresis loop of the spin source measured with a magnetic field applied along the stripes.   }
	\label{exp}
\end{figure*}

 In this letter, we present an experimental search for one of the interactions between electrons and nucleons described by 
\begin{equation}
V(r) =- f_{4+5} \frac{\hbar^2}{8\pi m_e c}[\hat{\sigma}\cdot(\vec{v}\times\hat{r}) ](\frac{1}{\lambda r}+\frac{1}{r^2})e^{-r/\lambda}, 
\label{eq1}
\end{equation} 
where $m_e$ and $\hat{\sigma}$ are the mass and the spin unit vector of the electron, respectively; $\vec{v}$ is the relative velocity between the electron and nucleon, and $r$ is the distance between them; $\lambda =\hbar/m_bc$ is the interaction range with $m_b$ being the mass of the exchanged bosons, and $f_{4+5}$ is the coupling constant; $\hbar$ is the reduced Planck constant, and $c$ is the speed of light in vacuum.  The interaction can be mediated by massive spin-1 bosons $Z'$ in the nonrelativistic limit, as described by the Lagrangian $\mathcal{L}_{Z'} = Z'_\mu\sum_{\psi}\bar{\psi}\gamma^\mu(g_V^\psi + \gamma_5g_A^\psi)\psi$. Here $\psi$ denotes the fermion field, and $f_{4+5}$ is equal to $\frac{1}{2} g_V^eg_V^N$\cite{Bogdan2006, Fadeev2019}.  Such interaction can also be generated by exchanging spin-0 bosons where  $f_{4+5}$ is equal to $\frac{1}{2} g_s^eg_s^N$. 

Various precision measurement technologies have been applied to search for this type of interaction\cite{Heckel2006, Heckel2008, Piegsa2012, Ficek2017, Ficek2018, Clayburn2023, Wu2022,xiao2023, Wu2022, Kim2018, Wu2023a, Wu2023, Ding2020}. Although no traces of the new interaction have been observed so far, increasingly stringent constraints have been imposed on its strength.  For the range above 100 m, Clayburn and Hunter set the strictest bounds by reanalyzing the data from the torsion-pendulum experiment using the  Earth as a rotating unpolarized source\cite{Clayburn2023, Wu2022}.  In the range of around 0.1 m, the most stringent constraints are imposed by atomic magnetometer experiments\cite{xiao2023, Wu2022, Kim2018}.  In the micrometer range, we had performed an experiment to set the strongest limits  on the interaction using a cantilever force sensor and an oscillating spin-modulated source \cite{Ding2020}. These constraints were improved by a recent experiment using solid-state spin sensors\cite{Wu2023a, Wu2023}. In this work, the constraints are further strengthened through several key experimental improvements.

The experimental scheme is similar to that previously reported\cite{Ding2020}. As shown in Fig. \ref{exp}(a), a soft cantilever is used to measure the force in the near-vertical direction between an unpolarized gold microsphere and a spatially modulated spin source. The signal of interest is modulated to a higher harmonic by applying a cosine voltage ($V_d$) to the piezoelectric stage to drive the source oscillating. A fiber optic interferometer is used to measure the deflection of the cantilever and then derive the force. The signal is demodulated numerically to obtain the real (Re) and imaginary (Im) parts at a specific harmonic using the displacement of the source as a phase reference. The oscillation of the source is monitored simultaneously with another fiber optic interferometer, as well as the drive voltage. 

In order to enhance the strength of the exotic interaction under a certain coupling constant, the following measures have been taken in this work. Since the interaction is proportional to the relative velocity, a higher oscillation frequency ($f_d$) and a larger amplitude ($A_d$) would produce greater exotic force. To this end, we have constructed a new scanning probe microscope with a compact and rigid design, which allows us to stably drive the source at a frequency of 18.85 Hz and an amplitude of 28.732 $\mu$m.  The oscillation amplitude is chosen to maximize the expected signal at the 10th harmonic. This effectively increases the relative velocity by a factor of $\sim$ 5.2 compared to Ref. \cite{Ding2020}. 

\begin{figure}
	\includegraphics {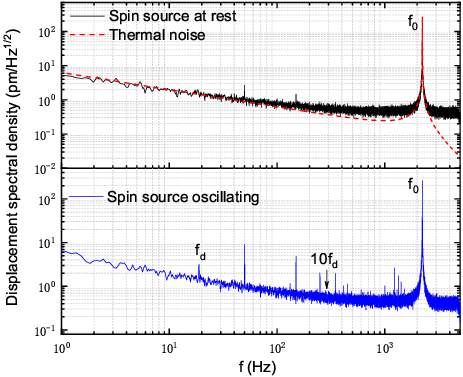}
	\caption{Cantilever displacement spectral density measured with and without the spin source oscillating. The thermal noise of the cantilever is also plotted. }
	\label{noise_floor}
\end{figure}

Further improvements have been adopted in the fabrication of the spin source. The spin source is made of alternative stripes of FeCo and SiO$_2$ [see Fig.\ref{exp} (c)] with a thickness of 705 nm. The reason to choose FeCo is that FeCo has a higher remnant magnetization than that of NiFe used in Ref. \cite{Ding2020}. In Fig. \ref{exp}(d), the magnetic measurements show that the spin source has a near rectangular hysteresis curve with a remnant magnetization being  94\% of the saturated magnetization. Together with the increase in thickness, the number of spins is increased to about 16 times. The fabrication process has also been optimized to reduce the thickness of the cover layer above the periodic structure, thereby reducing the separation between the gold sphere and the spin source.  Following the new fabrication process, we can fabricate spin sources with a gold cover layer of only 160 nm thick while keep the average periodic corrugation of the surface as low as 3 nm. 

The data are acquired with a commercial single-crystalline silicon cantilever glued with a gold microsphere at its end [shown in Fig.\ref{exp} (b)]. The cantilever spring constant is determined to be 33.1 mN/m by matching the simulated first and second resonance frequencies with the experimental measurements \cite{Wang2016}. Figure \ref{noise_floor} shows the cantilever displacement spectra measured at two states of motion of the source: at rest and oscillating at $f_d$ = 18.85 Hz at a far distance. The noise floor is at the same level for both spectra and no signal is observed at 10$f_d$, suggesting an increase in the mechanical stability of the new scanning probe microscope. The noise floor is close to the thermal noise of the cantilever and is 22 fN/$\sqrt{\rm Hz}$ at 10$f_d$. 

We demodulate the signal at 188.5 Hz, the 10th harmonic of the oscillation frequency, which is much larger than what is used in Ref. \cite{Ding2020}. For a oscillation of the source described with $x = x_0 +A_d {\rm cos} (2\pi f_d t)$, the expected signal at $10 f_d$ is given by 
\begin{equation}
\begin{aligned}
F^{ex}_{10}(x_0) = & -i \sum_{m=1}^\infty 2\pi f_d A_d {\rm Im} [f^{ex}(k_m)e^{ik_m x_0}][J_9(k_m A_d)
 \\ &+J_{11}(k_m A_d)],
\label{F45}
\end{aligned}
\end{equation} 
where $k_m = m2\pi /\Lambda$ and $\Lambda$ is the period of spin modulation, $f^{ex}(k_m)$ is the $m$th complex coefficient of the Fourier series expansion of $F^{ex}(x)/v$, and $F^{ex}(x)$ is the exotic force acting on the gold sphere. $J_9$ and $J_{11}$ are the Bessel functions of order 9 and 11, respectively. The $F^{ex}_{10}(x_0)$ is a periodic function of $x_0$ with a period of $\Lambda$. It is worth noting that the $F^{ex}_{10}(x_0)$ is purely imaginary for such a force that is proportional to the relative velocity. For a force $F^s(x)$ that depends only on the position, the signal is purely real as given by
\begin{equation}
F^s_{10f_d}(x_0) = - \sum_{m=1}^\infty J_{10}(k_m A_d) {\rm Re} [F^s(k_m)e^{ik_m x_0}],
\label{Fnormal}
\end{equation} 
where $F^s(k_m)$ is the $m$th complex coefficient of the Fourier series expansion of $F^s(x)$. These properties make the experiment intrinsically insensitive to the spurious forces commonly present in short-range experiments, such as the electrostatic force, Casimir force, and magnetostatic force.

\begin{figure*}
	\includegraphics{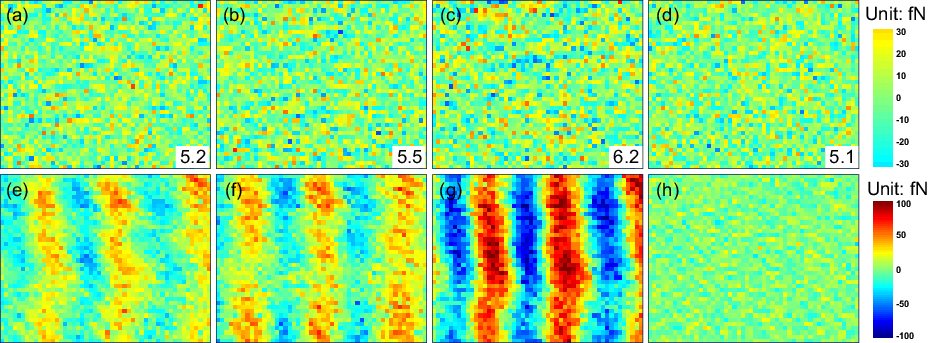}
	\caption{Images of the real (bottom) and imaginary part (top) of the signal. The size of the images is $40\times50$ $\mu$m$^2$. The values indicated at the corners of the images are the standard deviation of the entire image in fN. The residual potential difference with respect to the optimal compensation voltage is equal to (a), (e) 0, (b), (f) $-33$ mV, and (c), (g) 68 mV, respectively. (d), (h) Reference data taken at a far distance where the interaction between the sphere and spin source can be ignored. The top color scale bar is for (a), (b), (c), and (d), while the bottom color scale bar is for the rest of the images.}
	\label{2Dimages}
\end{figure*}

To search for the exotic force, the data are acquired on a 50$\times$40 grid of oscillation equilibrium positions ($x_0$, $y_0$) at a certain distance above the surface. A 22-second long time series data are taken at each grid point with a sampling frequency of 10 kHz. The data are then processed to demodulate numerically the real and imaginary parts of the signal at the 10th harmonic, resulting in images of the real and imaginary parts. Special attention needs to be paid to aligning the time series data with the displacement of the spin source in demodulation (See detailed discussion in End Matter). 

During data acquisition, the contact potential difference between the sphere and the spin source is compensated to minimize the electrostatic force. Prior to data acquisition, topographic images are taken with the same probe in the atomic force microscopy mode to verify that the source surface is free of significant defects in the area of data acquisition. The topographic images are also used to measure the inclination of the source surface for leveling  and tilt compensation.

\begin{table}
	\caption{Values and uncertainties of the main experimental parameters, and their contribution to the relative errors in the coupling constant estimation for $\lambda = 2.1$ $\mu$m.}
	\centering
	\begin{tabular*}{\hsize}{@{}@{\extracolsep{\fill}}lccc@{}}
		\hline\hline
		Parameter                               & Value                    & $\Delta f/ f$ 
		\\\hline
		Sphere diameter $2R$                 & $21.54\pm0.07$ $\mu$m    	          & 0.09\%              \\
		Period of the structure $\Lambda$     & $16.10\pm0.10$ $\mu$m                 & 0.66\%              \\
		FeCo thickness $h$                & $705\pm18$ nm                         & 2.1\%              \\
		FeCo width $W_1$      & $6.59\pm0.10$ $\mu$m                  & 0.19\%              \\
		FeCo width $W_2$      & $7.73\pm0.09$ $\mu$m                  & 0.13\%              \\
		FeCo spin density              & $(1.41\pm0.04)\times10^{29}$ m$^{-3}$ & 2.8\%              \\
		Distance $d$                      & $809\pm42$ nm                  & 2.6\%			   \\
		Drive amplitude $A_d$             & $28.732\pm0.008$ $\mu$m                   & $\textless$ 0.01\%        \\
		Spring constant $k$               & $33.1\pm1.6$ mN/m                     & 4.8\%              \\
		Int. Sensitivity $S$    & $102.4\pm0.3$ nm/V                         & 0.29\%              \\
		\hline\hline
		\label{parameter}
	\end{tabular*}
\end{table}

Figure \ref{2Dimages} shows typical images of the real and imaginary parts measured at a distance of 809 nm. The images of the imaginary part show a random background with no obvious signature of periodicity associated with the modulation structure. The standard deviation of the images is around 5.2 fN, which is close to the standard deviation of the reference data taken at a far distance, and is also close to the noise floor calculated from the displacement spectra with an integration time of 22 seconds. In contrast, the images of the real part show a periodic variation with the same period as the modulation structure. The amplitude of the variation increases with the increase of the residual potential difference after compensation, while the standard deviation of the imaginary images hardly changes. This observation is consistent with the expectation that the electrostatic force, as a velocity-independent force, contributes only to the real part of the signal, while the velocity-dependent exotic interaction contributes only to the imaginary part of the signal.

\begin{figure}
	\includegraphics[width=8cm] {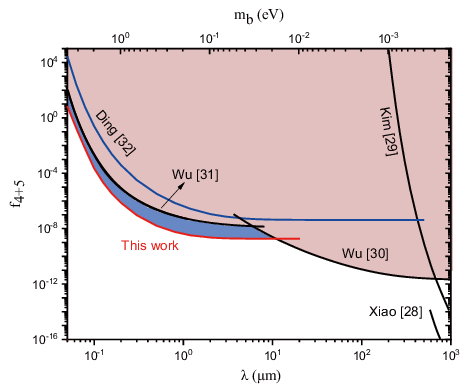}
	\caption{Constraints on the coupling constant of the exotic interaction. The solid red line represents the result of this work. The other lines are constraints reported in Ref. \cite{Ding2020,Wu2023,Wu2023a,Kim2018,xiao2023}.}
	\label{constraints}
\end{figure}

Since we observe no signature of the exotic interaction in the imaginary images, we set new constraints on this interaction in the following. We estimate the coupling constant for each $\lambda$ by fitting the predicted exotic force to the imaginary image using maximum likelihood estimation\cite{Ding2020, Wang2016, Ren2021}. The main experimental parameters and their contribution to the relative error of the coupling constant are listed in Table \ref{parameter}. Taking $\lambda = 2.1$ $\mu$m as an example, the fit yields a coupling constant $f_{4+5} = (-8.8\pm6.7)\times 10^{-10}$, implying a null result at the 95\% confidence interval. Figure \ref{constraints} shows the 95\% confidence exclusion bounds obtained by this work, as well as other limits reported in the literature\cite{Ding2020,Wu2023,Wu2023a,Kim2018,xiao2023}. We improve the limits for an interaction range below 10 $\mu$m, and the bound at $\lambda$ = 2.1 $\mu$m is about 10 times more stringent than the current limit.

In conclusion, we have performed an experimental search for the spin- and velocity-dependent exotic interaction using a cantilever force sensor. By increasing the relative velocity, the thickness and the spin density of the spin source, we are able to enhance the magnitude of the exotic force for a given coupling constant. Additionally, by accurately demodulating the signal at the 10th harmonic of the driving frequency, we successfully separate the signal of interest from spurious forces. These advancements allow us to strengthen the constraints on this interaction in the micrometer range. The measurements are currently limited by the  thermal noise of the cantilever. Further improvement can be made to reduce the electrostatic force and  then decrease the distance between the sphere and the spin source.

\begin{acknowledgments}
This work was supported by the National Key R\&D Program of China (Grant No. 2022YFC2204100) and the National Natural Science Foundation of China (NSFC) (Grant No. 12475052).
\end{acknowledgments}

\section{End Matter}
Signal demodulation: The data taken at every position are first filtered to remove the resonant signal of the cantilever.  Using the drive voltage as a reference, we  then  split the filtered data into segments with a length of $mT$, where $T$  is the oscillation period of the spin source and $m$ is an integer. The real and imaginary components are then calculated numerically for each segment, and their  mean and standard deviation are obtained for this position. Due to the nonlinear response of the piezoelectric stage to the drive voltage, the displacement of the spin source cannot be described by a simple cosine function even when we apply a drive voltage $V_d = V_0  {\rm cos}(2\pi f_d t)$. This results in a leak of the real component to the imaginary component, mimicking the exotic force signal.   In order to demodulate the signal accurately, we measure the displacement of the spin source with a fiber laser interferometer, and then fit the data to a Fourier series $x(t) = x_0 + \sum_{n=1} ^{10} A_n {\rm cos} (n\omega_d t + \phi_n)$ which produce much better fitting quality. A new time ($t'$) is then defined by setting $x(t) = x_0' + A_d {\rm cos}(2\pi f_d t')$. In terms of $t'$, the displacement is now a simple cosine function of time with an initial phase error less than 1 mrad.  At the end, we calculate the real and imaginary components with respect to $t'$ for every segment as follows 
\begin{equation}
	V_{\rm Re}(10f_d) =\frac{1}{mT}\int_{0}^{mT} V(t'){\rm cos}(20\pi f_dt'){\rm d}t'
	\label{FFF12}
\end{equation} 
\begin{equation}
	V_{\rm Im}(10f_d) = \frac{1}{mT}\int_{0}^{mT} V(t'){\rm sin}(20\pi f_dt'){\rm d}t' \\
	\label{FFF1}
\end{equation} 

%

\end{document}